\documentclass
[aps,preprint,twocolumn,superscriptaddress,tightenlines,10pt]{revtex4}%
\usepackage{amsmath}
\usepackage{color}
\usepackage{amsfonts}
\usepackage{amssymb}
\usepackage{graphicx}%
\begin{document}
\title{Tail-free self-accelerating solitons and vortices}
\author{Jieli Qin }
\affiliation{State Key Laboratory of Precision Spectroscopy, East China Normal University,
3663 North Zhongshan Road, Shanghai, China and Collaborative Innovation Center
of Extreme Optics, Shanxi University, Taiyuan, Shanxi 030006, China}
\affiliation{School of Physics and Electronics Engineering, Guangzhou University, 230
GuangZhou University City Outer Ring Road, Guangzhou, 510006, China, }
\author{Zhaoxin Liang}
\affiliation{Department of Physics, Zhejiang Normal University, Jinhua 321004, China}
\author{Boris A. Malomed}
\affiliation{Department of Physical Electronics, School of Electrical Engineering, Faculty
of Engineering, and Center for Light-Matter Interaction, Tel Aviv University,
Ramat Aviv 69978, Israel}
\author{Guangjiong Dong}
\affiliation{State Key Laboratory of Precision Spectroscopy, East China Normal University,
3663 North Zhongshan Road, Shanghai, China and Collaborative Innovation Center
of Extreme Optics, Shanxi University, Taiyuan, Shanxi 030006, China}

\begin{abstract}
Self-accelerating waves in conservative systems, which usually feature slowly
decaying tails, such as Airy waves, have drawn great interest in studies of
quantum and classical wave dynamics. They typically appear in linear media,
while nonlinearities tend to deform and eventually destroy them. We
demonstrate, by means of analytical and numerical methods, the existence of
robust one- and two-dimensional (1D and 2D) self-accelerating tailless
solitons and solitary vortices in a model of two-component Bose-Einstein
condensates, dressed by a microwave (MW) field, whose magnetic component
mediates long-range interaction between the matter-wave constituents, with the
feedback of the matter waves on the MW field taken into account. In
particular, self-accelerating 2D solitons may move along a curved trajectory
in the coordinate plane. The system may also include the spin-orbit coupling
between the components, leading to similar results for the self-acceleration.
The effect persists if the contact cubic nonlinearity is included. A similar
mechanism may generate 1D and 2D self-accelerating solitons in optical media
with thermal nonlinearity.

\end{abstract}
\maketitle

\section{Introduction}
Self-accelerating Airy waves were predicted in 1979 in the context of quantum
mechanics \cite{berry}. Then, this concept was transferred to optics
\cite{siv}, plasmonics \cite{plasmonics}, acoustics \cite{acc}, gas discharge
\cite{gas-discharge}, and hydrodynamics \cite{water}, using the similarity of
the linear Schr\"{o}dinger equation to the paraxial wave-propagation equation
in classical physics. These wave modes offer applications to plasma guiding
\cite{10}, signal transmission \cite{12,13}, laser-beam filamentation
\cite{14}, optical micromanipulation \cite{15,16,17,18,19,20}, generation of
"light bullets" \cite{21,22,23,24}, and so on \cite{26,27,28}.

Quantum self-accelerating waves have been experimentally demonstrated in
electron optics \cite{electron}. Self-accelerating Dirac waves have also been
predicted in relativistic quantum mechanics \cite{dirac}. Coherent
Bose-Einstein condensates (BEC) may be appropriate for the realization of the
self-acceleration in matter waves. The latter effect has not yet been
demonstrated experimentally, although it has been theoretically elaborated,
assuming the use of laser beams to imprint appropriate phase modulation onto
the BEC \cite{kli}, or the use of a trapping potential moving with
acceleration \cite{yuc}.

Ideal Airy waves with slowly decaying oscillatory tails carry an infinite
norm, therefore truncated Airy waves with a finite norm were used in the
theory and experiments \cite{siv,pan}; however, the truncation causes gradual
decay of the self-accelerating wave packets. Furthermore, the study of the
evolution of the Airy waves, which are eigenmodes of the linear propagation,
in various nonlinear media
\cite{ell,jia,abd,huy,ka,lot,fat,rud,pan,zhang,ka2,do,dri,all,Thawatchai}
shows that the nonlinearity causes deformation and, often, destruction of the
self-accelerating waves. Another type of self-accelerating solitary-wave pairs
was predicted \cite{Peschel} and experimentally demonstrated
\cite{Peschel-exper} in nonlinear photonic crystals with opposite signs of the
dispersion (effective mass) for the paired modes.

The above-mentioned settings were implemented neglecting dissipation in the
medium. On the other hand, robust optical tail-free self-accelerating pulses
have been predicted and experimentally demonstrated under the action of
various nonconservative effects, such as the sliding-frequency filtering
\cite{sliding}, ionization of the dielectric medium \cite{ionization},
diffusion in photorefractive crystals \cite{diffusion}, and intra-pulse
stimulated Raman scattering \cite{Raman}. In the latter case, the number of
photons (integral norm of the pulse) is conserved, but the Raman effect breaks
the conservation of the momentum and Hamiltonian.

The present work shows that the long-range nonlinear interaction between
constituents of a binary BEC, mediated by a microwave (MW) field (this
interaction was elaborated in Refs. \cite{we} and \cite{we2}), supports
spatially symmetric tail-free self-accelerating hybrid solitons, both one- and
two-dimensional (1D and 2D), built of matter-wave and MW components (in that
sense, they resemble exciton-polariton solitons, which are also matter-field
hybrids \cite{Exc}, although the present model is a strictly conservative one,
while exciton-polaritons states exist in dissipative semiconductor cavities,
therefore they should be supported by pump fields). It is relevant to stress
that both the self-trapping and self-acceleration are induced by the same
interaction, while in previously studied nonlinear systems the acceleration
was driven by terms such as the induced-Raman-scattering one, while the
self-trapping was provided by the Kerr nonlinearity. Further, we demonstrate
that the self-acceleration mechanism works equally well in the presence of the
spin-orbit coupling (SOC) between the two components of the BEC wave function,
as well as in the presence of the usual nonlinear contact interactions. In
addition to the results for the matter-wave solitons, it is demonstrated that
similar self-accelerating optical solitons can be produced in conservative
optical media with strongly nonlocal nonlinearity.

The rest of the paper is organized as follows. The 1D model is formulated in
Sec. II, which includes the linear SOC effect in the two-component BEC.
Systematic results, both analytical and numerical ones, for the
self-acceleration for 1D solitons are reported in Sec. III. The 2D extension
of the model is presented in Sec. IV, where emphasis is made on the stable
self-acceleration of vortices. This section also includes a brief
consideration of a nonlocal optical system which may be represented by a
similar model, thus predicting similar results for 1D and 2D self-accelerating
spatial optical solitons. The paper is concluded by Sec. V.

\section{The one-dimensional system}

\begin{figure}
[ptb]
\begin{center}
\includegraphics[
height=1.2532in,
width=2.5144in
]%
{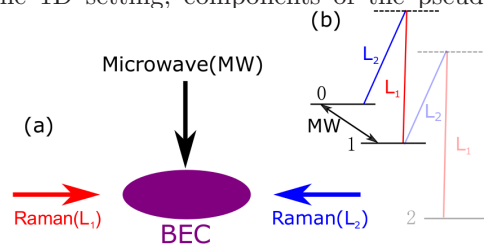}%
\caption{Counterpropagating Raman beams $L_1$ and $L_2$ generate the spin-orbit
coupled two-component BEC, as shown in Ref. \cite{Spielman}. The microwave
field (MW) couples states 0 and 1 by an effective long-range interaction, see
Eqs. (\ref{psi}) and (\ref{eq:green}). }%
\label{1}%
\end{center}
\end{figure}

Because some essential results reported below are obtained for the binary BEC
including the SOC effect, it is relevant to outline its implementation in the
relevant setting. It may be realized using the scheme shown in Fig. \ref{1}
\cite{Spielman}: counterpropagating Raman-laser beams $L_1$ and $L_2$ drive the
atomic gas, adiabatically eliminating level 2 and creating the SOC system with
pseudospin 1/2, whose components represent atoms in states 0 and 1. The
emulation of various aspects of gauge physics in ultracold gases by means of
SOC has drawn a great deal of interest \cite{lin}-\cite{liyun1}. In
particular, SOC solitons have been predicted in 1D \cite{1D,wubiao}, 2D
\cite{Ben-Li,chuanwei,2D-SOC-solitons}, and 3D \cite{3D} geometries, see also
a review in \cite{EPL}. However, SOC breaks the Galilean invariance, which
makes the creation of moving solitons a nontrivial issue
\cite{wubiao,Ben-Li,chuanwei}. It was found that 2D solitons in the system
with SOC of the Rashba type feature mobility only in one direction, up to a
critical value of the velocity, beyond which delocalization occurs
\cite{Ben-Li}. Self-accelerating solitons have not been found in previous
works dealing with SOC models.

We here consider the pseudo-spinor BEC with two components coupled by the
interaction mediated by the magnetic component of the MW field, as
schematically shown in Fig. \ref{1} \cite{we}. It has been found that the
interaction gives rise to quiescent hybrid MW-matter-wave solitons in 1D
\cite{we}, as well as to giant solitary vortices in 2D, which are stable, at
least, up to topological charge $S=5$ \cite{we2}. The solitons persist in the
presence of additional contact interactions of either sign, corresponding to
self- and cross-attraction or repulsion of the pseudo-spinor's components.%

In the 1D setting,\ components of the pseudo-spinor wave function, $\Psi
\equiv\binom{\Psi_{\downarrow}}{\Psi_{\uparrow}}$, which correspond to states
0 and 1 in Fig. \ref{1}, coupled by MW magnetic-field potential $H$, obey
coupled Gross-Pitaevskii equations (GPEs), which may or may not include the
SOC terms with strength $K$, represented by the first spatial derivatives,
which are combined with the Rabi-coupling frequency, $\Omega$:
\begin{widetext}
\begin{equation}
i\partial _{t}\left(
\begin{array}{c}
\Psi_{\downarrow} \\
\Psi_{\uparrow}
\end{array}
\right) =\left[ -\frac{1}{2}\partial _{x}^{2}+U\left( x\right)
+\left(
\begin{array}{cc}
iK\partial _{x} & \Omega -H \\
\Omega -H^{\ast } & -iK\partial _{x}%
\end{array}%
\right) -\left(
\begin{array}{cc}
\beta _{1} |\Psi_\downarrow|^2 + \beta _{2} |\Psi_\uparrow|^2 & 0 \\
0 & \beta _{1} |\Psi_\uparrow|^2 + \beta _{2} |\Psi_\downarrow|^2%
\end{array}%
\right) \right] \left(
\begin{array}{c}
\Psi_\downarrow\\
\Psi_\uparrow
\end{array}
\right).  \label{psi}
\end{equation}%
\end{widetext}Real coefficients $\beta_{1}$ and $\beta_{2}$ represent here,
severally, the self- and cross-component contact interactions, $\beta_{1,2}>0$
($<0$) corresponding, respectively, to the attractive (repulsive) sign of the
interactions. The wave function is subject to the normalization,
\begin{equation}
\int_{-\infty}^{+\infty}\Psi^{\dag}\Psi dx=1. \label{N}%
\end{equation}
The feedback of the matter-wave components on the MW potential (a specific
manifestation of the general \textit{local field effect} \cite{LFE}) is
accounted for by the Poisson equation \cite{we},
\begin{equation}
\partial_{x}^{2}H=-\gamma\Psi_{\downarrow}^{\ast}\Psi_{\uparrow},
\label{Poisson}%
\end{equation}
whose solution can be written with the help of the 1D Green's function:
\begin{equation}
H\left(  x,t\right)  =-\frac{\gamma}{2}\int_{-\infty}^{+\infty}\left\vert
x-x^{\prime}\right\vert \Psi_{\downarrow}^{\ast}\left(  x^{\prime},t\right)
\Psi_{\uparrow}\left(  x^{\prime},t\right)  dx^{\prime}. \label{eq:green}%
\end{equation}
Note that the asymptotic form of the potential, produced by Eq.
(\ref{eq:green}) at $|x|\rightarrow\infty$, is a linear function of the
coordinate, which is a commonly known property of solutions to the 1D Poisson
equation with a localized source of the field:%
\begin{equation}
H(x)\approx-\chi|x|,~\chi\equiv\left(  \gamma/2\right)  \int_{-\infty
}^{+\infty}\psi_{\downarrow}^{\ast}\left(  x\right)  \psi_{\uparrow}\left(
x\right)  dx. \label{x}%
\end{equation}

In the presence of SOC, Eq. (\ref{psi}), the position, energy and time are
scaled, respectively, by the inverse SOC wavenumber, $k^{-1}$, recoil
energy, $E_{R}\equiv\hbar^{2}k^{2}/(2m)$, and $\hbar/E_{R}$, so that
$K\equiv1$ is fixed in Eq. (\ref{psi}), unless we consider the system without
SOC, by setting $K=0$. Further, $U\left(  x\right)  $ in Eq. (\ref{psi}) is a
trapping potential (actually, we aim to consider free-space solitons, with
$U=0$), and
\begin{equation}
\gamma\equiv Nm\varepsilon_{0}\mu_{0}^{2}\omega_{\mathrm{MW}}^{2}M^{2}%
/(\hbar^{2}Ak^{3}) \label{gamma}%
\end{equation}
is the effective strength of the MW-mediated long-range interaction, with
$\varepsilon_{0}$ and $\mu_{0}$ being the vacuum permittivity and
permeability, $N$ the number of atoms, $m$ the atomic mass, $\omega
_{\mathrm{MW}}$ the MW frequency, $M$ the atomic magnetic moment, and $A$ the
confinement area in the transverse plane. Considering, for instance, BEC of
$^{87}$Rb atoms, transversely confined in area $\sim1$ $\mathrm{\mu}$m$^{2}$,
coupled to MW with wavelength of $\sim1$ mm, and the action of SOC with
wavenumber $k\sim 1 \mathrm{\mu} m^{-1}$, Eq. (\ref{gamma}) yields $\gamma
\sim10^{-9}N$. Thus, for BEC of $10^{7}$ atoms (actually, condensates made of
up to $10^{8}$ atoms are available, according to current experimental results
\cite{dis}), one obtains $\gamma\sim10^{-2}$. Following this estimate, we fix
$\gamma=0.02$ in numerical simulations following below. Furthermore, by using
even tighter transverse confinement to reduce the transverse-localization
area, $A$, the necessary number of atoms can be made essentially smaller than
$10^{7}$, which is used here as the estimate.

\section{Self-accelerating one-dimensional solitons}

First, it is necessary to produce stationary solitons with real chemical
potential $\mu$, in the form of%
\begin{equation}
\Psi\left(  x,t\right)  =\left(
\begin{array}
[c]{c}%
\psi_{\downarrow}(x)\\
\psi_{\uparrow}(x)
\end{array}
\right)  e^{-i\mu t},
\end{equation}
with the \textquotedblleft pseudo-spin-up" and \textquotedblleft down"
components of the complex spinor wave function, $\psi_{\uparrow,\downarrow
}(x)$, obeying the stationary free-space version of Eq. (\ref{psi}), with
$U(x)=0$:\begin{widetext}
\begin{equation}
\mu \left(
\begin{array}{c}
\psi _{\downarrow } \\
\psi _{\uparrow }%
\end{array}%
\right) =\left[ -\frac{1}{2}\partial_x^2 +\left(
\begin{array}{cc}
iK\partial_x & \Omega -H \\
\Omega -H^{\ast } & -iK\partial_x%
\end{array}%
\right) -\left(
\begin{array}{cc}
\beta _{1}|\psi_{\downarrow}|^2 + \beta _{2} |\psi_{\uparrow}|^2 & 0 \\
0 & \beta _{1}|\psi_\uparrow|^2 + \beta _{2} |\psi_{\downarrow}|^2%
\end{array}%
\right) \right] \left(
\begin{array}{c}
\psi _{\downarrow } \\
\psi _{\uparrow }%
\end{array}%
\right) .  \label{mu}
\end{equation}
\end{widetext}%
\begin{figure}
[ptb]
\begin{center}
\includegraphics[
height=2.5368in,
width=2.8747in
]%
{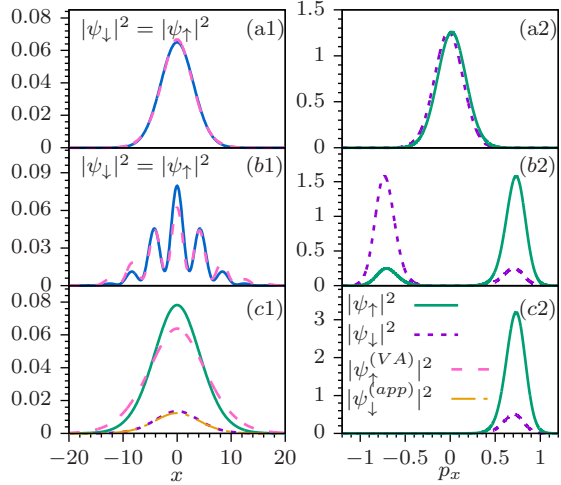}%
\caption{Typical density profiles in the coordinate space (left) and momentum
space (right, produced by the Fourier transform of the coordinate profile), of
a regular (single-peak) soliton (top), stripe soliton (middle), and a soliton
of the plane-wave type (bottom), so called because it is carried by the
plane-wave phase, for the pseudo-spin-up (solid lines) and spin-down
(short-dashed lines) components. These solutions to Eq. (\ref{psi}), with
$K\equiv1$ and $\beta_{1,2}=0$, were obtained, by means of the imaginary-time
integration, at $\Omega=1.5$, $\Omega=0.7$, and $\Omega=0.7$, with the
respective chemical potentials $\mu={-1.4779}$,{\ }${-0.7300}${, and }%
${-0}.7298$. The wave functions predicted by the VA (long-dashed lines),
$\psi_{\uparrow}^{\text{(VA)}}(x)$, based on the ansatz defined by Eqs.
(\ref{varw}) and (\ref{Ai}), with VA-predicted chemical potentials$-1.4781$%
,$-0.7262$ and $-0.7260$, respectively, are plotted too, for comparison with
the numerical results in the coordinate space. In panel (c1), the approximate
wave function $\psi_{\downarrow}^{\text{(app)}}$ for the spin-down component,
produced by the separately developed analytical approximation (\ref{app}), is
plotted by the yellow dotted-dashed line, which overlaps with its numerical
counterpart.}%
\label{so}%
\end{center}
\end{figure}

Actually, it is more convenient to produce the stationary wave function not
through Eq. (\ref{mu}), but rather solving Eqs. (\ref{psi})-(\ref{eq:green})
by dint of the imaginary-time-integration method. The simulations were carried
out in domain $|x|\leq80$, or in a smaller one, if it was sufficient for a
particular situation, with zero boundary conditions at edges of the domain. As
a result, three distinct species of the 1D solitons have been identified,
\textit{viz}., ones of the regular, stripe, and plane-wave types (solitons of
the latter type are carried by the plane-wave phase, see below). In the most
fundamental case, when the contact interactions are absent, i.e., $\beta
_{1,2}=0$ in Eq. (\ref{psi}), and solitons may only be supported by the
MW-mediated interaction, typical examples of the three species are displayed,
respectively, in the left top ($\Omega=1.5$), middle ($\Omega=0.7$), and
bottom ($\Omega=0.7$) panels of Fig. \ref{so} for $\gamma=0.02$, along with
the respective profiles in the momentum space, in the right panels. In
particular, the stripe and plane-wave-type solitons coexist at the same values
of parameters, being almost mutually degenerate under normalization condition
(\ref{N}), with chemical potentials $\mu={-0.7300}${\ and }${-0.7298}$,
respectively. A distinctive peculiarity of the soliton of the plane-wave type
is asymmetry between its components. {\LARGE \ }

The soliton of the regular type, presented in Fig. \ref{so}(a1,a2) for
$\Omega^{2}>1$, resembles those in the usual 1D SOC model (which does not
include the MW-mediated coupling between the components) \cite{review}, with
zero carrier wavenumber, $p_{x}=0$, while at $\Omega^{2}<1$\ the formal
linearization of Eq. (\ref{psi}) for tails of solitons [which implies setting
$H=0$, according to Eq. (\ref{eq:green})] demonstrates that they may develop
oscillations with wavenumbers $p_{x}=\pm\sqrt{\sqrt{\mu^{2}-\Omega^{2}}%
+1-|\mu|}$ (the solitons exist at $\mu<-1$). Solitons carried by $p_{x}$ with
the single sign are categorized as modes of the plane-wave type, while the
superposition of the two wavenumbers with opposite signs gives rise to stripe
solitons \cite{liyun1}. However, MW potential $H$, dressing the condensate,
completely changes the asymptotic shape (tails) of localized modes at large
$|x|$, where, taking Eq. (\ref{x}) into regard, Eq. (\ref{mu}) reduces to%
\begin{equation}
\mu\psi=\left(
\begin{array}
[c]{cc}%
-\frac{1}{2}\partial_{x}^{2}+iK\partial_{x} & \Omega+\chi|x|\\
\Omega+\chi^{\ast}|x| & -\frac{1}{2}\partial_{x}^{2}-iK\partial_{x}%
\end{array}
\right)  \psi, \label{v}%
\end{equation}
with $\chi$ defined as per Eq. (\ref{x}).

The presence of the effective linear potential $\sim|x|$\ in the asymptotic
equation (\ref{v}) suggests to approximate solutions by Airy functions
\cite{berry}. Accordingly, the variational approximation (VA) for solutions to
Eq. (\ref{mu}) may be based on the following ansatz:
\begin{equation}
\psi^{\mathrm{(VA)}}=\left[  \cos\alpha\binom{\cos\theta}{\sin\theta
}e^{-ip_{0}x}-\sin\alpha\binom{\sin\theta}{\cos\theta}e^{+ip_{0}x}\right]
\phi(x), \label{varw}%
\end{equation}
where the shape function is chosen as
\begin{equation}
\phi(x)=\psi_{0}{\mathrm{Ai(}}\left\vert x\right\vert /\sigma+\xi_{0}),
\label{Ai}%
\end{equation}
with $\xi_{0}\approx-1.019$ being the first local maximum of the Airy function,
${\mathrm{Ai}}(\xi)$, hence $x=0$ is the center of the adopted profile.
Variational parameters are $\alpha$, $\theta$, $p_{0}$, and $\sigma$, while
$\psi_{0}$ is a normalization constant. In contrast, in the usual 1D SOC
model, which does not include the MW field, $\phi(x)$ is approximated by a
sech or Gaussian ansatz \cite{liyun1}. At $\alpha=0$ or $\pi/2$, ansatz
(\ref{varw}) reduces to an envelope multiplying the plane wave, while at
$\alpha=\pi/4$, depending on $\Omega$, the ansatz may represent either a
stripe soliton, if wavelength $2\pi/p_{0}$ is small in comparison with the
envelope's width, or a regular single-peak soliton otherwise. Values of the
variational parameters are numerically determined by minimizing the system's
energy,
\begin{gather}
E=\int_{-\infty}^{+\infty}dx\psi^{\dagger}\left[  -\frac{1}{2}\frac{d^{2}%
}{dx^{2}}+\left(
\begin{array}
[c]{cc}%
iK\partial_{x} & \Omega\\
\Omega & -iK\partial_{x}%
\end{array}
\right)  \right]  \psi\nonumber\\
+\frac{\gamma}{4}\int\int dxdx^{\prime}\left\vert x-x^{\prime}\right\vert
\left[  \psi_{\downarrow}^{\ast}\left(  x\right)  \psi_{\uparrow}\left(
x\right)  \psi_{\downarrow}^{\ast}\left(  x^{\prime}\right)  \psi_{\uparrow
}\left(  x^{\prime}\right)  +{\mathrm{c.c.}}\right]  . \label{energy}%
\end{gather}
In Eq. (\ref{energy}), the contact interactions are again disregarded, setting
$\beta_{1,2}=0$ in Eq. (\ref{psi}), aiming to address the fundamental setting,
with the two components of the pseudo-spinor wave function interacting solely
via the MW magnetic field. The VA-predicted soliton profiles are displayed in
Fig. \ref{so} along with their numerically found counterparts, demonstrating
that the VA is reasonably accurate, providing, in particular, an accurate
approximation for the regular solitons.

\begin{figure}
[ptb]
\begin{center}
\includegraphics[
trim=0.007986in 0.039850in -0.007986in -0.039850in,
height=1.9906in,
width=3.0034in
]%
{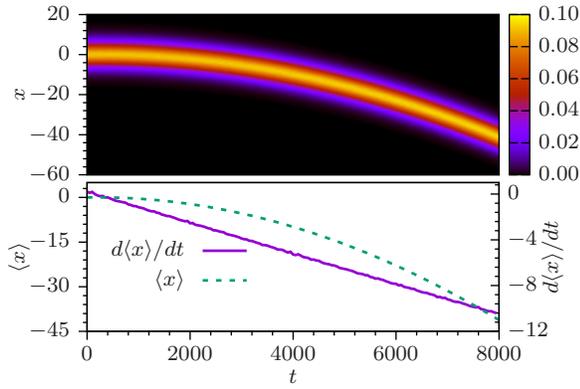}%
\caption{Top: The evolution of the same plane-wave-type soliton as in Fig.
\ref{so} (i.e., in the absence of the contact interactions, $\beta_{1,2}=0$),
initiated by the addition of a smal random perturbation to it. In this figure
and similar ones displayed below, the evolution is displayed by means of the
map of the total density of both matter-wave components in the $\left(
x,t\right)  $ plane. Bottom: the soliton's average position, $\left\langle
x(t)\right\rangle \equiv\int_{-\infty}^{+\infty}\Psi^{\dag}(x)\Psi(x)xdx$, and
velocity, $d\left\langle x(t)\right\rangle /dt$, as functions of time. The
time dependence of the velocity helps to evaluate the soliton's acceleration
(weak jitter in $d\left\langle x\right\rangle /dt$\ is caused by the
randomness of the initial perturbation).}%
\label{pw}%
\end{center}
\end{figure}

The MW-mediated long-range interaction, which is the underlying ingredient of
the present system, determines not only the shape of the solitons, but also
their dynamics, as shown by systematic simulations of Eq. (\ref{psi}). It was
found that the addition of small random perturbations to the regular and
stripe solitons does not produce any conspicuous effect (which implies that
they are completely stable modes in the quiescent state), while perturbed
solitons of the plane-wave type start self-accelerated motion, keeping their
integrity, as shown in Fig. \ref{pw} for the system without the contact
interactions ($\beta_{1,2}=0$).%

\begin{figure}
[ptb]
\begin{center}
\includegraphics[
trim=0.007986in 0.039850in -0.007986in -0.039850in,
height=1.9897in,
width=3.0025in
]%
{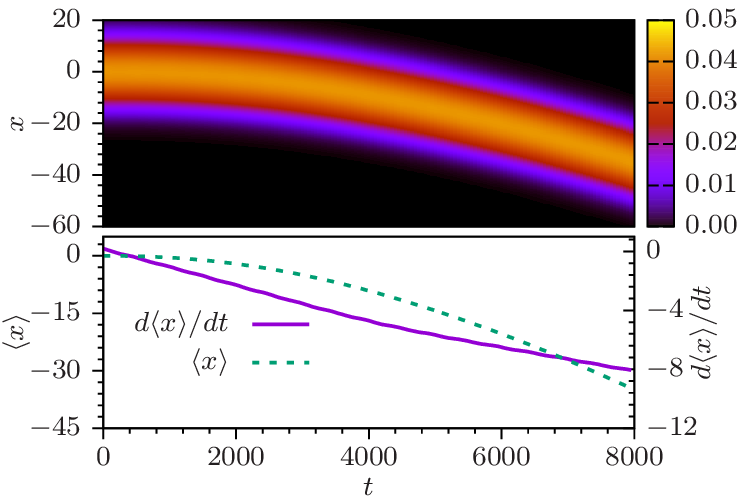}%
\caption{The same as in Fig. \ref{pw}, but in the presence of the
self-repulsive contact interactions in Eq. (\ref{psi}) with $\beta_{1}=-5.0$,
$\beta_{2}=0$.}%
\label{repulsive}%
\end{center}
\end{figure}

\begin{figure}
[ptb]
\begin{center}
\includegraphics[
trim=0.007986in 0.039850in -0.007986in -0.039850in,
height=1.9904in,
width=3.0024in
]%
{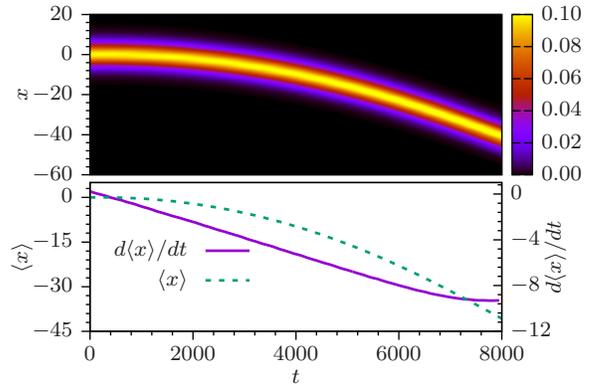}%
\caption{The same as in Fig. \ref{pw}, but in the presence of the
self-attractive contact interactions in Eq. (\ref{psi}), with $\beta_{1}=0.3$,
$\beta_{2}=0$.}%
\label{attractive}%
\end{center}
\end{figure}

Stable self-accelerating solitons persist in the presence of the contact
nonlinearity with the repulsive sign, as shown in Figs. \ref{repulsive}, as
well as under the action of relatively weak local attraction, see Fig.
\ref{attractive}. Strong attraction may change the situation, as it tends to
transform the solitons considered here, characterized by the ansatz based on
Eqs. (\ref{varw}) and (\ref{Ai}), into usual sech-shaped solitons, for which
the interaction with the MW field becomes negligible. Naturally, the strong
self-repulsion makes the moving soliton much broader, as seen in Fig.
\ref{repulsive}. As concerns the gradual decrease of the acceleration,
observed in Fig. \ref{repulsive}, starting from $t\simeq4500$, and in Fig.
\ref{attractive}, starting from $t\simeq7000$, detailed consideration of the
numerical data demonstrates that this effect is explained by a brake force,
which is applied to the soliton by radiation emitted by it at the initial
stage of the evolution, and eventually reflected from the edge of the
integration domain.

These results suggest existence of a family of stable self-accelerating
solitons of the plane-wave type, including the quiescent and moving solitons
displayed in Figs. \ref{so}(c1,c2) and (\ref{pw}), respectively, the
acceleration being an internal parameter of the family. In the system
dominated by the SOC terms, the family can be constructed in an analytical
form. To this end, we look for the corresponding solution to Eq. (\ref{psi})
as $\Psi_{\uparrow\downarrow}\left(  x,t\right)  =\operatorname{exp}\left(
ip_{0}x-i\mu t\right)  \phi_{\uparrow\downarrow}\left(  x,t\right)  $, with
slowly varying amplitudes $\phi_{\uparrow\downarrow}\left(  x,t\right)  $,
carrier momentum $p_{0}$, and chemical potential $\mu$. Using Eq. (\ref{psi}),
the small component [$\phi_{\downarrow}$, see Fig. \ref{so}(c1)] is eliminated
in favor of the larger one:
\begin{equation}
\phi_{\downarrow}(x)\approx C\left(  H^{\ast}-\Omega\right)  \phi_{\uparrow
}(x), \label{app}%
\end{equation}
\ with $C\equiv\left(  p_{0}-\mu+p_{0}^{2}/2\right)  ^{-1}$ [recall we set
$K\equiv1$ in Eq. (\ref{psi}), if the SOC terms are present]. This
approximation for $\phi_{\downarrow}(x)$ is plotted in Fig. \ref{so}(c1) by
means of the yellow dashed-dotted curve, showing very good agreement with its
numerically found counterpart. Further, substituting Eq. (\ref{app}) in the
remaining equation for $\phi_{\uparrow}$ in system (\ref{psi}) leads, in the
first approximation \cite{note0}, to a single GPE, in which term $CH^{\ast
}\phi_{\uparrow}$ from Eq. (\ref{app}) couples wave function $\phi_{\uparrow}$
to potential $H$:
\begin{equation}
i\partial_{t}\phi_{\uparrow}=\left[  -\frac{1}{2}\partial_{x}^{2}%
-i(p_{0}-1)\partial_{x}+2C\Omega H-\beta_{1}\left\vert \phi_{\uparrow
}\right\vert ^{2}\right]  \phi_{\uparrow}, \label{single}%
\end{equation}
Further, the substitution of relation (\ref{app}) in Eq. (\ref{Poisson}) leads
to the Poisson equation, which gives rise to real MW potential $H$ [unlike the
complex field, produced by Eqs. (\ref{Poisson}) and (\ref{eq:green}) in the
general case]:%
\begin{equation}
\partial_{x}^{2}H=C\gamma\Omega|\phi_{\uparrow}|^{2}. \label{H}%
\end{equation}

Then, the existence of self-accelerating solitons is explained by the fact
that Eqs. (\ref{single}) and (\ref{H}) are invariant with respect to an
\emph{exact transformation} from the laboratory reference frame to one moving
at an arbitrary constant acceleration, $a$ (while the usual GPE, as well as
nonlinearly coupled GPE systems, are not invariant with respect to this
transformation \cite{Quebec,Parker}):
\begin{gather}
\phi_{\uparrow}=\phi_{\uparrow}^{\prime}\left(  x^{\prime},t\right)
\operatorname{exp}\left[  i\left(  axt+\frac{1-p_{0}}{2}Ct^{2}-\frac
{a^{2}t^{3}}{6}\right)  \right]  ,\nonumber\\
x^{\prime}=x-\left(  a/2\right)  t^{2},~H^{\prime}=H+(2C\Omega)^{-1}ax.
\label{Phi'}%
\end{gather}
Thus, any quiescent soliton produced by Eqs. (\ref{single}) and (\ref{H})
generates a family of solitons moving with arbitrary acceleration ($a$), which
is indeed the intrinsic parameter of the family, as conjectured above. In
particular, term $(2C\Omega)^{-1}ax$, added to the MW potential by the
transformation, naturally breaks the symmetry of the linear in $x$ asymptotic
field (\ref{x}) with respect to to $x>0$ and $x<0$.

Note that, when $p_{0}=1$\ [i.e., the SOC term vanishes in Eq. (\ref{single}%
)], transformation (\ref{Phi'}) remain valid, and Eqs. (\ref{single}) and
(\ref{H}) are tantamount to the model introduced in Ref. \cite{we} without SOC
(however, moving solitons were not considered in that work). Thus, the
existence of the family of robust self-accelerating solitons does not depend
on the presence of the SOC terms in Eq. (\ref{psi}), and it is not broken
either by the inclusion of the contact-interaction term with coefficient
$-\beta_{1}$. The invariance with respect to the self-accelerating
transformation cannot be produced in an exact form for the full system of Eqs.
(\ref{psi}) and (\ref{Poisson}), but its ability to maintain the
self-acceleration is clearly demonstrated by numerical results (in particular,
by Figs. \ref{pw}-\ref{attractive}).

The GPE system may also be reduced to the single equation in the case opposite
to that considered above, namely, if the Rabi coupling dominates over SOC in
Eq. (\ref{psi}), i.e., $\Omega$\ is a large parameter. In this case, the
substitution of
\begin{align}
\Psi_{\uparrow}(x,t)  &  =\operatorname{exp}\left[  ix+\left(  1/2-\mu
_{0}\right)  it\right]  \Phi_{\uparrow}\left(  x,t\right)  ,\nonumber\\
\Psi_{\downarrow}\left(  x,t\right)   &  =\operatorname{exp}\left[
-ix+\left(  1/2-\mu_{0}\right)  it\right]  \Phi_{\downarrow}\left(
x,t\right)  , \label{PhiPhi}%
\end{align}
\ with $\mu_{0}=\pm\Omega$\ and slowly varying amplitudes $\Phi_{\uparrow
\downarrow}\left(  x,t\right)  $, yields a relation between them,
\begin{equation}
\Phi_{\downarrow}\approx\mu_{0}^{-1}\left(  \Omega-H^{\ast}\right)
e^{2ix}\Phi_{\uparrow}. \label{Phi}%
\end{equation}
On the contrary to the above case, when component $\phi_{\downarrow}$ was
small in comparison with $\phi_{\uparrow}$, as per Eq. (\ref{app}), Eq.
(\ref{Phi}) implies that the absolute values of the two components are nearly
equal. Eventually, the substitution of expressions (\ref{PhiPhi}) and
(\ref{Phi}) in Eqs. (\ref{phi}) and (\ref{Poisson}) leads to the following
equations, which differ from Eq. (\ref{single}), with $p_{0}=1$ (without the
SOC term) and Eq. (\ref{H}) only by the notation for coefficients:
\begin{gather*}
i\partial_{t}\Phi_{\uparrow}=\left[  -(1/2)\partial_{x}^{2}\mp2H(x)-\left(
\beta_{1}+\beta_{2}\right)  \left\vert \Phi_{\uparrow}\right\vert ^{2}\right]
\Phi_{1},\\
\partial_{x}^{2}H=\mp\gamma\left\vert \Phi_{\uparrow}(x^{\prime})\right\vert
^{2},
\end{gather*}
i.e., the single GPE limit is a \emph{universal }one, being equally relevant
in the cases of strongly unequal and nearly equal components of the
pseudo-spinor wave function.

\section{Two-dimensional systems}

\subsection{The pseudo-spinor condensate coupled by the microwave field}

In the 2D setting, Eqs. (\ref{psi}) and (\ref{Poisson}) are replaced by
equations which combine the 2D version of SOC
\cite{Ben-Li,chuanwei,2D-SOC-solitons} and the interaction of the
pseudo-spinor wave function with the MW field in two dimensions \cite{we2}:
\begin{widetext}
\begin{eqnarray}
i\partial _{t} \left(\begin{array}{c}
\Psi_\downarrow \\
\Psi_\uparrow
\end{array}\right) & = & \left[ -\frac{1}{2}\left( \partial _{x}^{2}+\partial
_{y}^{2}\right) +\left(
\begin{array}{cc}
iK\partial _{x} & K\partial _{y}+\Omega -H \\
-K\partial _{y}+\Omega -H^{\ast } & -iK\partial _{x}%
\end{array}%
\right) \right]\left(\begin{array}{c}
\Psi_\downarrow \\
\Psi_\uparrow
\end{array}\right)  \\ \nonumber
& &-\left(
\begin{array}{cc}
\beta _{1}|\Psi_\downarrow|^2 + \beta _{2}|\Psi_\uparrow|^2 & 0 \\
0 & \beta _{1}|\Psi_\uparrow|^2 + \beta _{1}|\Psi_\downarrow|^2%
\end{array}%
\right) \left(\begin{array}{c}
\Psi_\downarrow \\
\Psi_\uparrow
\end{array}\right) ,  \label{psi2D}
\end{eqnarray}%
\end{widetext}
\begin{equation}
\left(  \partial_{x}^{2}+\partial_{y}^{2}\right)  H=-\gamma\Psi_{\downarrow
}^{\ast}\Psi_{\uparrow}. \label{Poisson2D}%
\end{equation}
To provide straightforward insight into dynamics of the 2D system, we again
resort to the limit case of the strong Rabi coupling between the two
components in Eq. (\ref{psi2D}), which dominates over SOC. Then, two
components of the pseudo-spinor may be reduced to one, cf. Eq. (\ref{Phi}),
and the system of Eqs. (\ref{psi2D}) and (\ref{Poisson2D}) amounts to the
single-component GPE coupled to the 2D Poisson equation, i.e., as a matter of
fact, the 2D extension of Eqs. (\ref{single}) and (\ref{H}):
\begin{equation}
i\partial_{t}\Phi_{\uparrow}=-\frac{1}{2}\left(  \partial_{xx}^{2}%
+\partial_{yy}^{2}\right)  \Phi_{\uparrow}\mp2{\mathrm{Re}}(H)\Phi_{\uparrow
}-\left(  \beta_{1}+\beta_{2}\right)  |\Phi_{\uparrow}|^{2}\Phi_{\uparrow},
\label{Phi-2D}%
\end{equation}%
\begin{equation}
\left(  \partial_{xx}^{2}+\partial_{yy}^{2}\right)  H=\mp\gamma\left\vert
\Phi_{\uparrow}\right\vert ^{2}. \label{H-2D}%
\end{equation}

A straightforward but crucially important fact is that, similar to what was
found above for the 1D system [see Eq. (\ref{Phi'})], Eqs. (\ref{Phi-2D}) and
(\ref{H-2D}) are invariant with respect to the transformation to the reference
frame which moves in the 2D space with arbitrary initial velocities $V_{x}%
$,$~V_{y}$ and arbitrary constant accelerations $a_{x}$,$~a_{y}$:
\begin{gather}
x^{\prime}=x-V_{x}t-(1/2)a_{x}t^{2},~y^{\prime}=y-V_{y}t-(1/2)a_{y}%
t^{2}~,\nonumber\\
H=H^{\prime}\pm(a_{x}x+a_{y}y)/2,\nonumber\\
\Phi_{\uparrow}=\Phi_{\uparrow}^{\prime}\left(  x^{\prime},y^{\prime
},t\right)  \operatorname{exp}\{i[a_{x}t+V_{x})x+(a_{y}t+V_{y})y-\phi
(t)]\},\nonumber\\
\phi(t)=\frac{\left(  a_{x}t+V_{x}\right)  ^{3}-V_{x}^{3}}{6a_{x}}%
+\frac{\left(  a_{y}t+V_{x}t\right)  ^{3}-V_{y}^{3}}{6a_{y}}. \label{phi}%
\end{gather}
Accordingly, coordinates $\left(  x_{c},y_{c}\right)  $ of the center of the
stable 2D soliton (which may carry embedded vorticity \cite{we2}) move as
$x_{c}=V_{x}t+(1/2)a_{x}t^{2},~y_{c}=V_{y}t+(1/2)a_{y}t^{2}$, which may be a
curvilinear trajectory in the 2D plane: at small $t$, it is close to a
straight line with slope $x/y=V_{x}/V_{y}$, while at $t\rightarrow\infty$ it
becomes asymptotically close to a line with $x/y=a_{x}/a_{y}$. In particular,
in the case of $a_{x}=V_{y}=0$ the trajectory is a parabola: $y_{c}=\left(
a_{y}/2V_{x}^{2}\right)  x_{c}^{2}$.

Note that the solution of the 2D Poisson equation (\ref{H-2D}) has the
standard logarithmic asymptotic form far from the region where the source of
the field is located:%
\begin{equation}
H\approx\mp\frac{\gamma}{2\pi}\left(  \int\int\left\vert \Phi_{\uparrow
}\left(  x,y\right)  \right\vert ^{2}dxdy\right)  \ln r. \label{ln}%
\end{equation}
The difference of the field component of the self-accelerating 2D solitons
from that of their quiescent counterparts is more essential than in the 1D
case, as the potential terms linear in $x$ and $y$ [see Eq. (\ref{phi})] are
qualitatively different from the logarithmic term in Eq. (\ref{ln}).

\begin{figure}
[ptb]
\begin{center}
\includegraphics[
height=2.9301in,
width=2.9748in
]%
{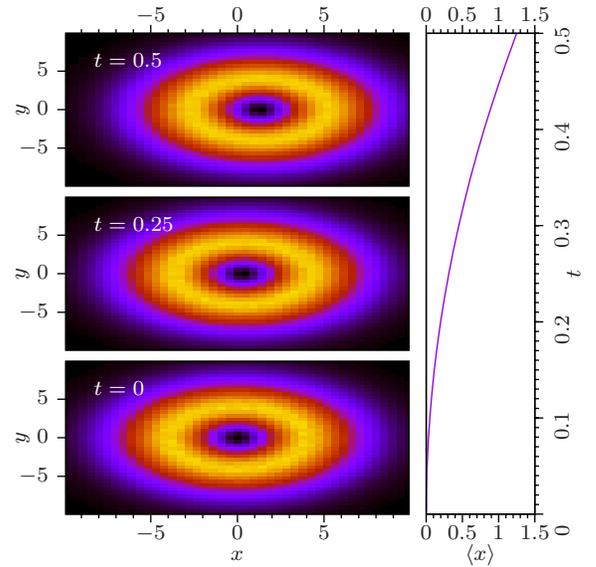}%
\caption{{$|\Phi_{\downarrow}|^{2}=|\Phi_{\uparrow}|^{2}$ for a} stable
self-accelerating vortex soliton with {$a_{x}=10$, $a_{y}=0$, $V_{x,y}=0$,
$\beta_{1,2}=0$ [no contact interactions in Eq. (\ref{Phi-2D})], $\gamma=\pi$%
}. The right panel shows the time evolution of coordinate $x$ of the vortex'
center.}%
\label{vor_beta=0}%
\end{center}
\end{figure}

\begin{figure}
[ptb]
\begin{center}
\includegraphics[
height=2.9301in,
width=2.9748in
]%
{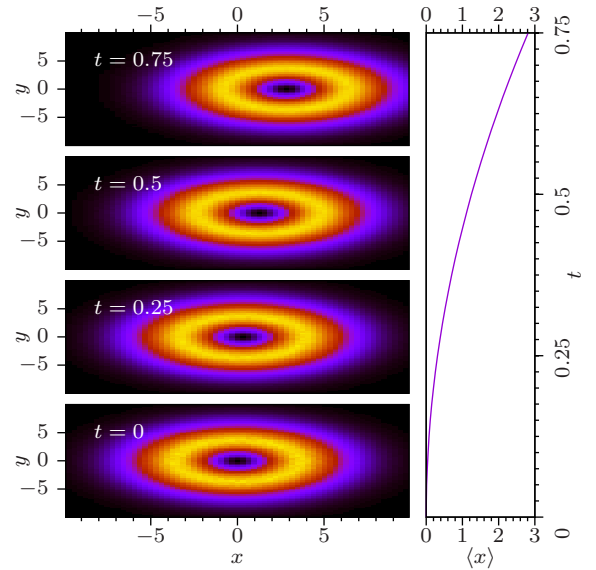}%
\caption{{The same as in Fig. \ref{vor_beta=0}, but in the presence of
relatively strong contact self-attraction, represented by coefficient }%
$\beta_{1}+\beta_{2}=10$ in Eq. (\ref{Phi-2D}).}%
\label{vor}%
\end{center}
\end{figure}

The predictions are corroborated, in Figs. \ref{vor_beta=0} and \ref{vor}, by
numerical solutions of Eqs. (\ref{Phi-2D}) and (\ref{H-2D}) (in the absence
and presence of the contact interaction, severally) for stable 2D solitons
with embedded vorticity $S=1$ (vortex rings), which move at a constant
acceleration in the $x$ direction, in exact agreement with Eq. (\ref{phi}). A
remarkable fact is that, in the case shown in Fig. \ref{vor}, the accelerating
vortex soliton remains stable in the presence of a relatively strong contact
self-attraction with $\beta_{1}+\beta_{2}=10$ in Eq. (\ref{Phi-2D}), in spite
of the well-known propensity of the cubic self-attraction to destabilize 2D
vortex-ring solitons against the collapse and ring splitting \cite{2D}. Note
also that the action of the self-attraction naturally leads to compression of
the vortex ring. Finally, it is relevant to stress that the acceleration
observed in Figs. \ref{vor_beta=0} and \ref{vor} is much larger than in Figs.
\ref{pw}-\ref{attractive}, because in the latter case it was induced by small
random perturbations initially added to the 1D solitons, while in the
situation displayed in Figs. \ref{vor_beta=0} and \ref{vor} the acceleration
was explicitly added to the input generating the 2D vortex solitons, as per
Eq. (\ref{phi}).
%

\subsection{The nonlocal optical system}

The mechanism of the formation of the tailless self-accelerating solitons,
elaborated above in terms of the two-component BEC coupled by the MW field,
can be realized as well in a nonlocal optical model with amplitude $E\left(
x,y,z\right)  $ of the electromagnetic wave and local perturbation $n\left(
x,y,z\right)  $ of the refractive index, which are governed by the coupled
system of the paraxial propagation equation and an equation which determines
how the index perturbation is created by the field distribution
\cite{nonlocal}:
\begin{gather}
iE_{z}+(1/2)\left(  E_{xx}+E_{yy}\right)  +nE=0,\label{E}\\
n-l^{2}\left(  n_{xx}+n_{yy}\right)  =|E|^{2}. \label{n}%
\end{gather}
Here $z$ is the propagation distance, $x$ and $y$ are transverse coordinates,
and $l$ is the correlation length of the nonlocality. Rescaling $E\equiv
l{\mathcal{E}}$ and taking the limit of a strong nonlocality, $n/l\rightarrow
0$, Eqs. (\ref{E}) and (\ref{n}) are reduced to a form tantamount to
Eqs.\ (\ref{Phi-2D}) and (\ref{H-2D}):
\begin{gather}
i{\mathcal{E}}_{z}+(1/2)\left(  {\mathcal{E}}_{xx}+{\mathcal{E}}_{yy}\right)
+n{\mathcal{E}}=0,\nonumber\\
n_{xx}+n_{yy}=-|{\mathcal{E}}|^{2}. \label{opt}%
\end{gather}
The 1D reduction of the 2D system (\ref{opt}) is obviously possible too. Thus,
stable 1D and 2D self-accelerating solitons may be predicted in this optical
setting too.

\section{Conclusion}

The objective of this work is to investigate the dynamics of the binary BEC
whose components, representing different hyperfine atomic states, are coupled
by the magnetic component of the MW (microwave) field. In the general case,
the SOC (spin-orbit coupling) is included too. The effective interaction
between the two components via the feedback of the atomic states on the MW
supports self-trapped modes (solitons), whose asymptotic form is the same as
that of the Airy function, in the 1D case. In the presence of SOC, we have
found stable 1D solitons of the regular (single-peak) and stripe types, and
solitons in the form of envelopes carried by plane waves. The most essential
finding is the existence of stable self-accelerating solitons of the
plane-wave type and their 2D counterparts, including vortex solitons. In
contrast to the previously studied self-accelerating Airy waves \cite{siv}, in
the present system the solitons keep simple self-trapped shapes, without
oscillatory tails attached to them, hence their integral norm is well defined
and convergent, unlike the divergent norm of the exact Airy waves. The present
system, being conservative, is also different from previously studied models
which admit self-acceleration of localized modes in optical media featuring
nonconservative effects, such as the sliding filtering, diffusion, ionization,
and stimulated Raman scattering. The existence of the family of the
self-accelerating solitons is demonstrated analytically, by reducing the
two-component systems, in the 1D and 2D setting alike, to a single GPE,
coupled to the Poisson equation for the potential of the MW field. These
reduced systems admit the exact transformation to a reference frame moving
with arbitrary acceleration, thus generating self-accelerating solitons from
quiescent ones in the exact form. In the 2D geometry, the transformation
generates 2D solitons which may move along curved trajectories, due to the
interplay between the 2D velocity and acceleration. The self-acceleration
mechanism persists if the contact nonlinearity is included. It can also be
realized in strongly nonlocal optical media, with the 1D or 2D transverse geometry.

\textbf{Acknowledgments} G.J. Dong acknowledges the support by the National
Science Foundation of China (grants No. 11574085 and 91536218), and 111
Project ( B12024),  the National Key Research and Development Program of China
(Grant No. 2017YFA0304201)，as well as Innovation Program of Shanghai
Municipal Education Commission. Z.X.Liang acknowledges the support of
the\ National Science Foundation of China (Grant No. 11374125), the key
projects of the Natural Science Foundation of China (Grant No. 11835011) and
Youth Innovation Promotion Association of the Chinese Academy of Sciences
(Grant No. 2013125). J. L. Qin acknowledges the support by the National
Science Foundation of China (No. 11847059). The work of B.A.M. was supported,
in part, by the joint program in physics between NSF and Binational
(US-Israel) Science Foundation through project No. 2015616, and by the Israel
Science Foundation through grant No. 1286/17. This author appreciates
hospitality of the State Key Laboratory of Precision Spectroscopy at East
China Normal University (Shanghai).


\begin{thebibliography}{99}
\bibitem {note0}A shift of the chemical potential, $p_{0}^{2}/2-p_{0}%
-C\Omega^{2}-\mu$, and a term $\sim|H|^{2}$ are neglected here, as they are
small in comparison with the term kept in Eq. (\ref{single}).
\end{thebibliography}
\end{document}